# Ultra-High-Temperature Vacuum Prober for Electrical and Thermal Measurements

Laurent Jalabert,[1,2,*], Jose Ordonez-Miranda,[1,2,3], Yunhui Wu,[2] Byunggi Kim[2,4], Roman Anufriev[1,2], Masahiro Nomura[2,1] and Sebastian Volz.[1,2]

[1] LIMMS, CNRS-IIS IRL 2820, The University of Tokyo, Tokyo, 153-8505, Japan
[2] Institute of Industrial Science, The University of Tokyo, Tokyo, 153-8505, Japan
[3] Sorbonne Université, CNRS, Institut des Nanosciences de Paris, INSP, F-75005 Paris, France
[4] Department of Mechanical Engineering, Institute of Science Tokyo, Tokyo, 152-8550, Japan

We develop an ultra-high-temperature vacuum probe station (UHT-VPS) featuring a sample holder heated by the thermal radiation from a silicon carbide heater. This contactless configuration electrically isolates the sample from the high-power heating source through a vacuum gap, ensuring reliable measurements under extreme condition. The capability of this UHT-VPS to measure electrical signals from 30 nV upward on bulk sapphire is demonstrated through the 3ω/2ω method. The measurements are continuously operated from 300 to 1150 K, under a high vacuum of $2 \times 10^{-6}$ mbar, for a total of about 66 hours without readjusting the contact, and yield the linear and quadratic temperature coefficients of resistance of chromium/platinum micro-resistances, as well as the sapphire's thermal conductivity and thermal diffusivity. By recording the heater and sensor temperature signals up to 30 kHz and fitting them with theoretical models that take into account the quadratic TCR of Cr/Pt microwires, we obtain values in agreement with literature data determined by optical methods. In this range of temperature, we also measure thermal conductivity, which cannot be directly accessed by optical methods. Our system thus provides an effective solution for simultaneously retrieving the electrical and thermal properties of materials using a single set of 3ω/2ω data up to unprecedented temperature levels.

## I. INTRODUCTION

High temperatures (up to 900 K) have dramatic deteriorating effects on silicon-based devices[1–3], including the increase of the intrinsic carrier concentration, a reduction of carrier mobility, changes in the crystal structure, uncontrolled dopant diffusion, and reduced reliability due to excessive leakage currents. Some issues can be overcome using wide band-gap semiconductors[4–6] (i. e. diamond, silicon carbide, gallium nitride, aluminum nitride), leading to a new generation of transistors[7,8] and microsystems[9–12] able to routinely operate even in harsh environments at high temperatures. In addition to those wide band gap materials, ultra-high temperature (UHT) ceramics[13], carbon-based devices[14–16], non-volatile memories (NVM)[17,18] or printed RF devices[19] have a wide variety of high temperature applications in energy harvesting, solar concentrators, nuclear plants as well as in automotive, aeronautics, ground transportation, space exploration[20], manufacturing, geothermal energy, deep-well-drilling, petroleum and military industries.

Ultra-high temperatures are routinely achieved by probe-less techniques especially for the thermal characterization of materials (thermal conductivity and diffusivity) conducted through different methods like the hot wire[21,22], the hot disk[23], the laser flash[24–33], the thermo-reflectance techniques[34,35], or modulated photothermal radiometry[36]. Thermal measurements at extreme temperature (> 2000 K) have been reported using laser flash method[28].

Yet, in electrical engineering—where a variety of materials are involved, and wire bonding or mechanical contacts with the sample pads are typically required—such extreme-temperatures remain a significant challenge for both electrical measurements and packaging[37]. With the new generation of devices operating beyond 1000 K[7,20], the ability to test and select the best devices from wafer-scale fabrication becomes more critical. Currently, the accelerated aging protocol defined by the standardized

*Corresponding author: jalabert@iis.u-tokyo.ac.jp

High Temperature Operation Lifetime (HTOL), is limited to 125-300°C and rely on Arrhenius-based lifetime prediction models[38].

Therefore, ICs and MEMS industries require novel characterization tools that can operate above 1000 K, enabling continuous wafer-level electrical testing over extended periods before dicing and packaging the most reliable devices.

In that way, Vacuum Probe Stations (VPS) are employed in semiconductor backend processing and academic research for electrical characterization, where measurements are carried out through mechanical contact between movable probe tips and the contact pads of a selected device on the wafer. These measurements are routinely performed with commercial VPS, at sample holder temperatures ranging from 4 to 800 K under vacuum conditions to prevent convection and corrosion.

Although the existence of numerous refractory materials capable of generating extreme temperatures in vacuum environments—including tungsten, molybdenum, tantalum, and graphite—the maximum operational holder temperatures reported in the literature for electrical measurements in prototypes of VPS were 773 K[39] with boralectric heater (with an upper temperature limited by the RF probes degradation), 923 K with a boron nitride heater (without micromanipulators and without testing a device)[40], 973 K (using fixed probes)[41], and recently long-term retention data was demonstrated in NVM at 1073 K in a compact VPS for small samples, using a SiC heater[18].

The implementation of refractory metal heaters does not appear as a fundamental limitation for VPS. Rather, the main challenge lies in managing the excessive thermal loads[18] imposed on the surrounding materials, such as ceramics, electrical probes, sensors, and the vacuum chamber itself. At elevated temperature, stainless-steel components can exhibit enhanced crack propagation and accelerated thermal fatigue degradation[42,43], requiring inner radiation shielding and/or advanced cooling engineering.

Alternative to probe stations have been reported on single device, using spring probes contact and mainly for thermoelectricity measurements[44–46], demonstrating that temperature limitations stem primarily from probe station design constraints (with movable probe tips for sequential testing of several devices on a wafer), rather than fundamental measurement physics (which can be done on a single device, without the need to probe on a wafer).

In this work, we address the challenge of continuous electrical measurements at ultra-high temperature by developing a VPS able to operate from 300 to 1150 K for several hours, with a cooling engineering of the stainless-steel chamber to prevent damages and for a safe manipulation. To illustrate a wide range of operability, we perform measurements via the electro-thermal 3ω method[47–50] involving high stability, low-level voltages, long-time measurements on a wide variety of device (bulk, thin films, superlattices, suspended membranes, phononic crystals, surface phonon polaritons, …). Recent improvements lead to the 3ω/2ω[51,52] method with extended capabilities to measure the in-plane[53] and cross-plane thermal conductivity and thermal diffusivity of solid materials[51], including the possible radiative losses[54]. After determining the linear and non-linear TCR of chromium/platinum micro wires, we accurately retrieve both the thermal conductivity and thermal diffusivity of bulk sapphire by fitting the theoretical model to the recorded 3ω and 2ω temperature signals[51]. These automated high-temperature measurements go beyond the state of the art, as the highest temperature reported for the 3ω method is 780 K[55].

## II. PROBER DESIGN

Figure 1(a) shows the developed UHT-VPS equipped with six probes mounted on micromanipulators. A silicon carbide (SiC) heater (Fig.1(b)) is connected to a high-power DC source. The SiC heater is placed a few millimeters below a molybdenum holder (supporting the sample), so that the sample temperature is lower than the SiC heater. By avoiding physical contact between the high-power SiC heater and the device, thermal radiation heating prevents electrical leakage (See supplementary material Fig. S1), eliminates the need for ceramic insulation, and ensures safe and robust isolation from surrounding metal components. An input power of 478 W yields a uniform holder temperature of 1150 K (Fig. 1(c) and Fig. S2).

*Corresponding author: jalabert@iis.u-tokyo.ac.jp

While SiC heaters could operate at extreme-temperatures, their durability might be limited by oxidation-related degradation[56]. Similarly, standard K-type thermocouples, made of Chromel and Alumel, can degrade due to oxidation and contamination from the sheath[57]. Therefore, a high-vacuum condition ($2 \times 10^{-6}$ mbar, Pfeiffer HiCube300) is needed, not only to prevent convection losses, but also to enhance the durability of thermal components by minimizing oxidation.

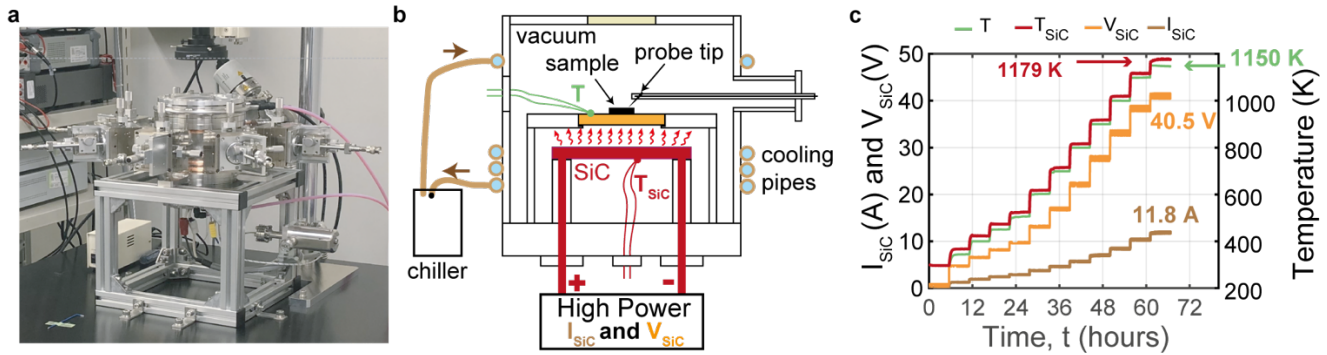

**Figure 1.** (a) High-temperature VPS implemented with six probes (SMA connectors) connected to a vacuum pump (not shown) and a chiller. (b) schematic of the system with a DC high power source plugged to a silicon carbide heater, heating the top plate by radiation. (c) The top plate and SiC temperatures are recorded with thermocouples type K (TC). The maximum temperature was 1179 K on the SiC heater (1150 K on the top plate) for an input power of 478 W.

Since non-contact heating leads to a long-time lag, the standard Proportional-Integral-Derivative (PID) algorithm is not suitable, especially for controlling temperatures below 600 K. Therefore, we developed an ad-hoc temperature control based on an experimental calibration (top plate thermocouple versus the DC supplied voltage) using the software LabVIEW (National Instrument) with a feedback control on two type-K thermocouples. One of these thermocouples is attached to the SiC heater ($T_{SiC}$) and another one is arc-welded to the top plate (T), as close as possible to the sample, preventing the use of a sheath. The device is mechanically clamped on the holder. Six electrically isolated probe tips can contact the pads of a device under test, and are connected to shielded SMA connectors on the micro-positioner. The gold-coated tungsten tips can move ±5 mm with a step of 5 μm, which allows moving the tips and contact another device on the wafer. A chiller (Eyela CA1116-A) circulates a liquid coolant at 300 K in the surrounding areas of the top plate, the chamber walls, and the probe arms preserving the stainless-steel from degradation, and ensuring safe manipulation even when during the SiC heater operation at 1179 K.

## III. SAMPLE FABRICATION

We applied our UHT-VPS to determine the thermal properties of monocrystalline sapphire bulk using the 3ω/2ω method. A 0.5-mm-thick sample of single crystal sapphire (Shinkosha Co., Ltd, α-Al₂O₃ of 10 × 10 mm², orientation [0001] c-plane, double-side polished) was chosen for its high performance in power electronics, high melting point (≈2320 K), and good chemical properties that make it suitable for operating in harsh environments and temperatures above 1000 K.

*Corresponding author: jalabert@iis.u-tokyo.ac.jp

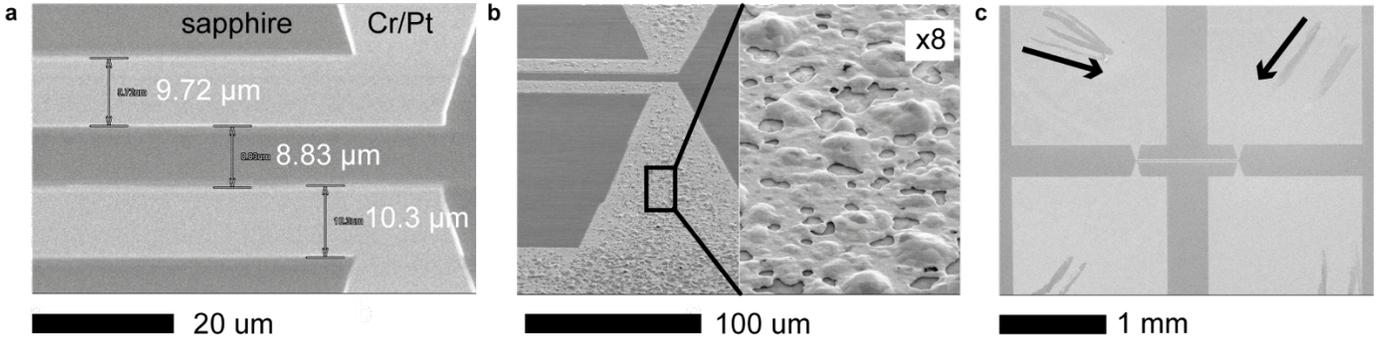

**Figure 2.** Scanning electron microscopy image of the Cr/Pt micro-resistances (a) as-deposited on a sapphire sample and (b) after the experiment at 1150 K (insight magnification x8 from the left image). The 200-nm-thick platinum thin film exhibits islands at random locations. (c) Thermal expansion of the probe arms featuring scratches of about 700 μm along the black arrows.

To evaluate the cross-plane and in-plane thermal properties of our sapphire sample, we deposited two symmetric resistive Cr/Pt micro-wires on the surface using photolithography, DC sputtering (Anelva, E-200S, 0.6 Pa, 40 W, 10 sccm Argon gas) and a lift-off process. Each Cr/Pt wire is typically 10 μm in width, 0.20 μm in height (See Fig. S3), and 1000 μm in length. The wires are separated by a distance of 8.83 μm, as shown in Fig. 2(a).

The sapphire surface roughness between the two microwires is below 1.2 nm (See Fig. S3). The process was optimized to ensure a high-quality interface between the metal and the substrate and prevent the peeling of the thin metallic films at ultra-high temperatures. Our thermal measurements were limited to 1150 K because of the degradation of the 0.20 μm thick Cr/Pt layer patterned on the sapphire sample (Fig. 2(b)), which is independent of the performance of the VPS. From the initial platinum thin film (Fig. 2(a)), the high-temperature experiment induced metal depletions and island formations (Fig. 2(b) and Fig. S3) which are consistent with in-situ reported observations[12,58]. Refractory metals (melting point >2000ºC) such as molybdenum[44] may be considered as an alternative to the platinum thin film. Here, the maximum temperature is low enough to prevent the formation of terraces in the sapphire c-plane[59]. The micro heater current density was $0.78 \times 10^6$ A/cm$^2$, and might have slightly accelerated the wire degradation by electromigration.

The oversized pads with wide triangular extensions were designed to minimize their contribution to the total resistance (which should be concentrated in the narrow resistive wire), and accommodate long probe drift due to the unavoidable thermal expansion of the probe arms. Figure 2(c) shows several scratches of about 700 μm in length, left by each probe drift on each pad during the continuous 66-hour experiment (See supplementary material video_UHT_VPS.mp4), but without requiring any readjustment of the probe-pad contacts.

## IV. THE 3ω/2ω SETUP

The scheme of our automated 3ω/2ω setup is shown in Fig. 3. The heater side consists of a classic AC half-bridge with a programmable resistance box ($R_{hp}$) embedding analog foil resistors and set to 160 Ω. Each of these resistors has a TCR $< 5 \times 10^{-6}$ K$^{-1}$. Low-noise differential amplifiers (Analog Device, AD524) were used to reduce the parasitic 1ω signal and detect the 3ω signal (Fig4(a)) with a dual frequency lock in amplifier. Due to the wide range of the holder temperature, for a constant injected current $i_{1\omega}$, the micro resistance $R_h$ increases by a factor three and the voltage can bypass the limit of 1 V at the input of standard lock in amplifier. Typically, the micro heater and sensor resistances were around 89 Ω and reached 265 Ω at 1150 K as shown on Fig. 5 (and Fig. S4). Therefore, we implement a dual frequency and high-input lock in amplifier for measuring the voltage $V_\omega$ up to a maximum of 10 V (Fig. S5). The amplitude of the AC heater current $i_{1\omega}$ was set to 15.5 mA (11 mA rms) to ensure a reliable 2ω sensor signal detection at the highest frequency.

*Corresponding author: jalabert@iis.u-tokyo.ac.jp

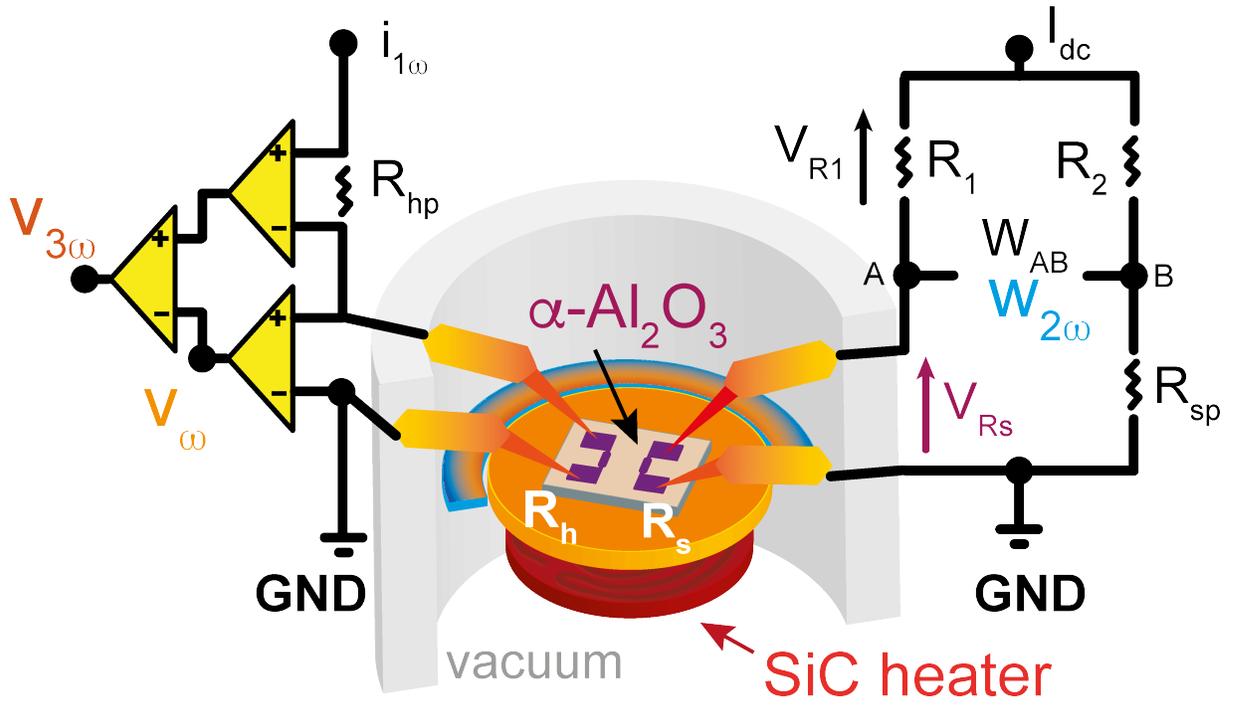

**Figure 3.** Scheme of the electrical 3ω/2ω setup implemented with the sample and its patterned resistors (violet wires) loaded in the VPS. The micro heater/sensor is connected to a half/full Wheatstone bridge.

On the sensor side, we implemented a DC Wheatstone bridge aiming at reducing the DC parasitic voltage and measuring the sensor signals in the nanovolt range. The fixed metal foil resistances $R_1 = R_2 = 50$ Ω have a $TCR = 2.5 \times 10^{-6}$ K$^{-1}$. The variable external resistance $R_{sp}$ is set by a programmable resistance box, in which each analog resistor has a $TCR < 5 \times 10^{-6}$ K$^{-1}$. To ensure the DC voltage exceeded the noise level of the digital multimeters (DMM), a low-noise DC current source $I_{dc} = 1$ mA was applied to the Wheatstone bridge (resulting in less than 50 µW of DC power dissipated in $R_s$). Although this induces some micro sensor self-heating, it has no impact on the 2ω voltage measured by the lock-in amplifier. The 2ω component of the sensor voltage contributes less than 0.05 % to the total DC voltage $V_{Rs}$, and is therefore considered to be negligible (see Fig. 4(b) and (d)).

By circulating an AC current at frequency ω along the micro heater, thermal waves appear and propagate at 2ω from the heater to the sensor. On the sensor side, another lock-in amplifier measures the bridge differential voltage at the second harmonic 2ω (Fig 4(b) and Fig. S6). A digital multimeter measures the DC voltage drop $V_{Rs}$ (Fig. 4(d)) to evaluate the sensor micro resistance $R_s$, after knowing the DC voltage $V_{R1}$ measured across the fixed resistance $R_1$ and hence the current flowing in $R_s$. Such a direct method was found to be as accurate (< 0.15 %) (See Table S1) as retrieving $R_s$ from the DC bridge voltage between the points A and B shown in Fig. 3.

Both the heater/sensor were powered with low noise AC/DC current sources, which allow changing the modulation frequency and amplitude of the supplied electrical current without discontinuity of the output signal. A LabVIEW program controls the settings of eleven instruments, adjusts the instruments sensitivity from a feedback control on the input signals, and continuously records the data with a typical sampling time of three seconds (3 s). The position of the probe tips was unchanged during a total duration of 66 hours. The stabilization time is 2.2 hours for the holder temperature (with a sampling time of 40 s) and ten minutes for the heater current or frequency change (with a sampling time of 3 s). The program automates a set of experiments based on sweeping the holder temperature from 300 to 1150 K and the heater frequency from 30 kHz down to 0.71 Hz for each temperature.

*Corresponding author: jalabert@iis.u-tokyo.ac.jp

The highest frequency (30 kHz) was set below the attenuation frequency $f_c = 33.8$ kHz given by Eq. (18b), where $D = 10.63$ mm$^2$/s is the measured thermal diffusivity of the sapphire substrate at 296 K (See Table S4) and $a = 5$ μm is the heater half-width.

The micro resistances $R_h$ and $R_s$ were calculated from the time-dependent voltages shown in Figs. 4(b-c). For each plateau of temperature and frequency, the heater 1ω voltage $V_\omega$ (Fig. 4(c)) and the sensor DC voltage $V_{Rs}$ (Fig. 4(d)) were stable with time, except at the highest temperature of 1150 K, for which a drastic increase was observed due to the platinum damage shown in Fig. 2(b). All measurements were taken continuously, without readjusting the contact between the tips and the pads.

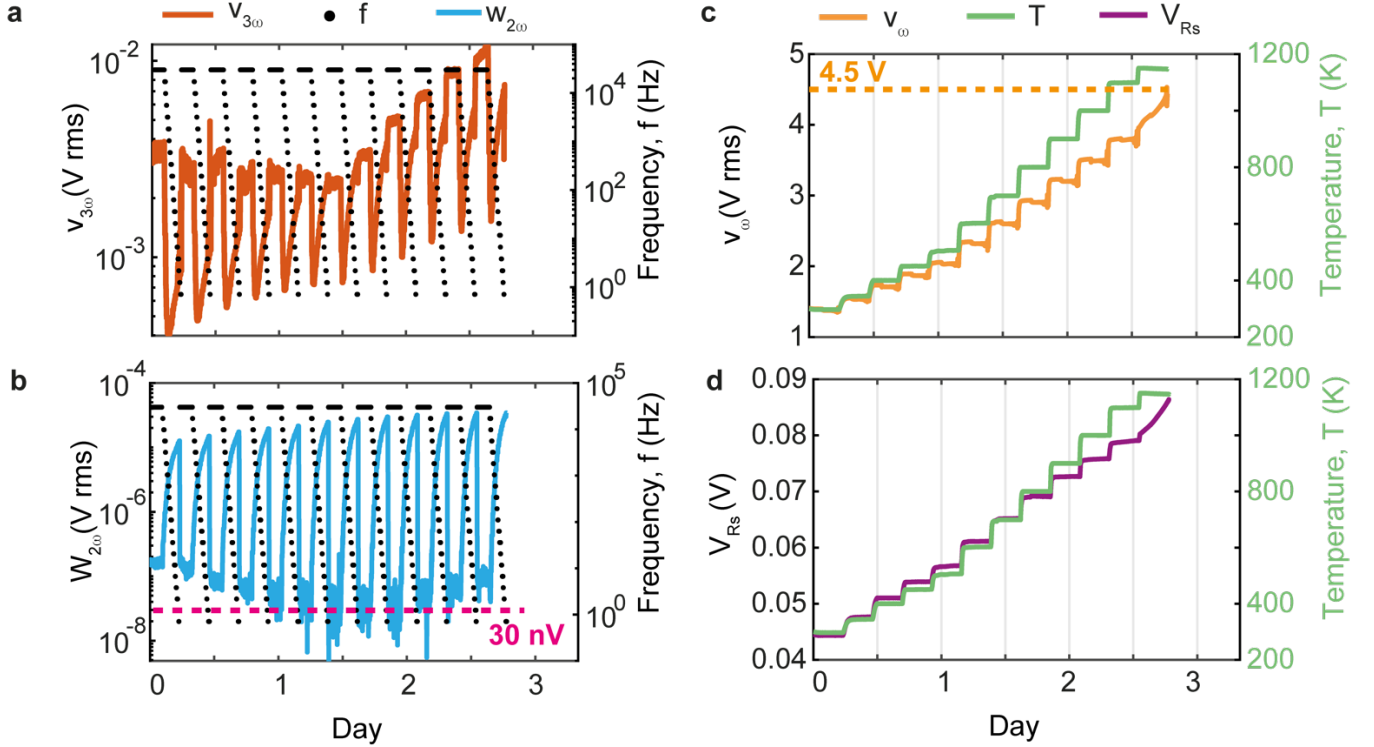

**Figure 4.** (a) The heater 3ω voltage and (b) the sensor 2ω voltage in the Wheatstone bridge showing the minimum amplitude of 30 nV (green dashed line), (c) AC voltage measured across the micro heater with a high input lock-in amplifier and the (d) DC voltage measured across the micro sensor with a digital multimeter, for a holder temperature $T$ swept from 300 to 1150 K and a frequency scan from 30 kHz to 0.71 Hz. At 1150 K, the voltages $V_\omega$ and $V_{Rs}$ increase due to the damage of the Pt thin film.

## V. TEMPERATURE COEFFICIENT OF RESISTANCE (TCR)

The conversion of electrical data to thermal variables, such as the local elevation of temperature, relies on the accuracy of determination of the temperature coefficient of resistance (TCR) of both the micro heater and sensor. Since the TCR depends mostly on the deposition technique, the thickness, and pre-annealing of the thin film[12], the sample was annealed in vacuum at 1150 K for about one hour prior to the measurement, in an attempt to stabilize the intrinsic properties of the thin film.

*Corresponding author: jalabert@iis.u-tokyo.ac.jp

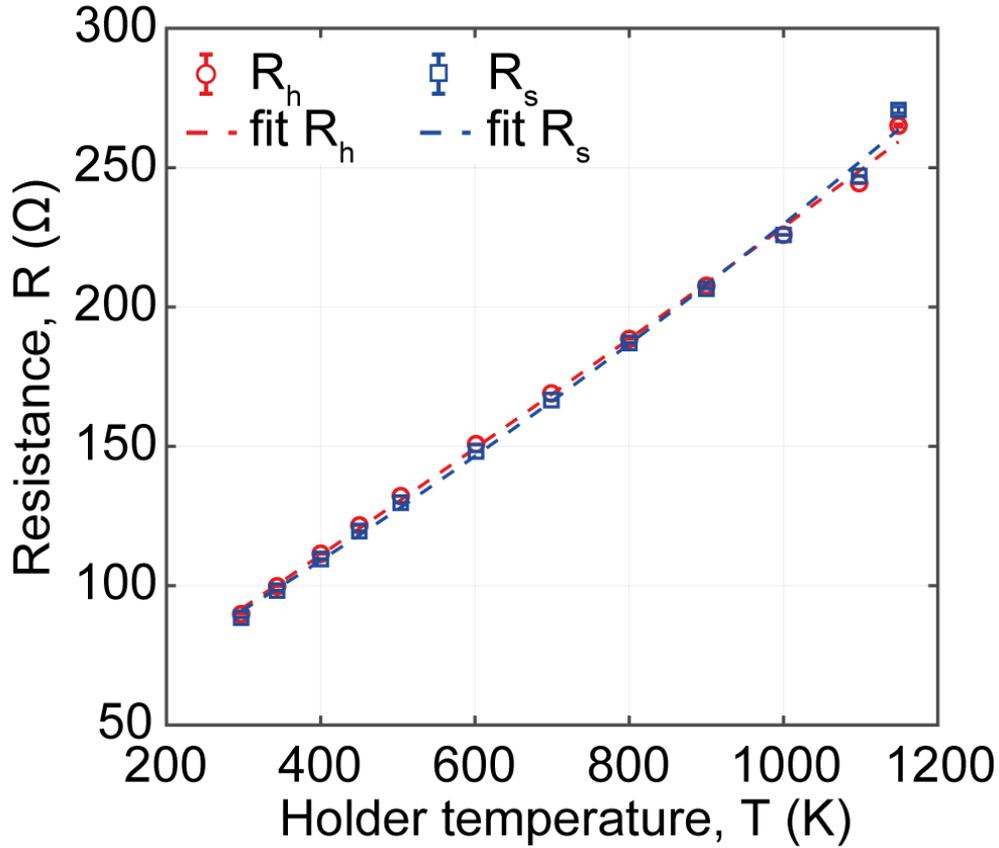

**Figure 5.** Heater and sensor Cr/Pt micro resistances (200 nm-thick) as functions of the holder temperature and for a frequency of 30 kHz. The fittings (dashed lines) were done with Matlab nonlinear regression model (Eq. 2) in the machine learning toolbox, for retrieving the linear ($\alpha$) and quadratic ($\beta$) temperature coefficients of resistance (TCRs) and their standard deviations (See Table S1-S2).

To calculate the TCRs, the mean and standard deviation of all the signals were computed from the last 20% of the recorded data (i.e., the final 120 seconds of a 600-second acquisition for each frequency and holder temperature; an example is given in Fig. S7).

The resulting heater ($R_h$) and sensor ($R_s$) resistances data were taken for each holder temperature at the highest measurement frequency (30 kHz), which corresponds to the smallest 3$\omega$ and 2$\omega$ signal amplitudes, thence the smallest temperature rise in both elements (See Fig. S8).

The values of the mean and standard deviation of the heater $R_h$ and sensor $R_s$ resistances are shown in Fig. 5, as functions of the holder temperature. On the heater [sensor] side, the linear $\alpha$ and quadratic $\beta$ TCRs of $(2.036 \pm 0.069) \times 10^{-3}$ K$^{-1}$ [$(1.903 \pm 0.085) \times 10^{-3}$ K$^{-1}$)] and $(1.300 \pm 0.976) \times 10^{-7}$ K$^{-2}$ [$(3.868 \pm 0.120) \times 10^{-7}$ K$^{-2}$) were respectively obtained by nonlinear regression fitting with a correlation degree $R^2 = 0.998$ [0.997] (See Table S2 and Fig. S9). Those TCRs are consistent with data reported in the literature[12,60–64] (See Table S3). As needed for sensing, the linear TCR values of the micro resistances are about 400 times higher than the TCR of the fixed or programmable external resistors. As a result of the non-linear temperature evolution of these micro resistances from 300 to 1150 K, we extended the standard 3$\omega$/2$\omega$ modeling to capture the effect of the quadratic TCR of each micro-resistance.

## VI. MODELING OF THE ELECTRIC SIGNALS

*Corresponding author: jalabert@iis.u-tokyo.ac.jp

Here we model the electrical voltages read by the heater and sensor on top of an anisotropic substrate, as shown in Fig. 6. The circulation of an electrical current $I = I_0\cos(\omega t)$ modulated in time $t$ with a frequency $\omega$ along the heater, generates an electrical power $P(t) = R_0 I^2 = 0.5 R_0 I_0^2[1 + \cos(2\omega t)]$, with $R_0$ being its electrical resistance at the sample (holder) temperature $T_0$. The power amplitude $P_0 = 0.5 R_0 I_0^2$ is about 10.85 mW. The electrical power injected to the heater and converted into heat via Joule effect thus splits into a steady-state component and a temporal one oscillating with a frequency $2\omega$.

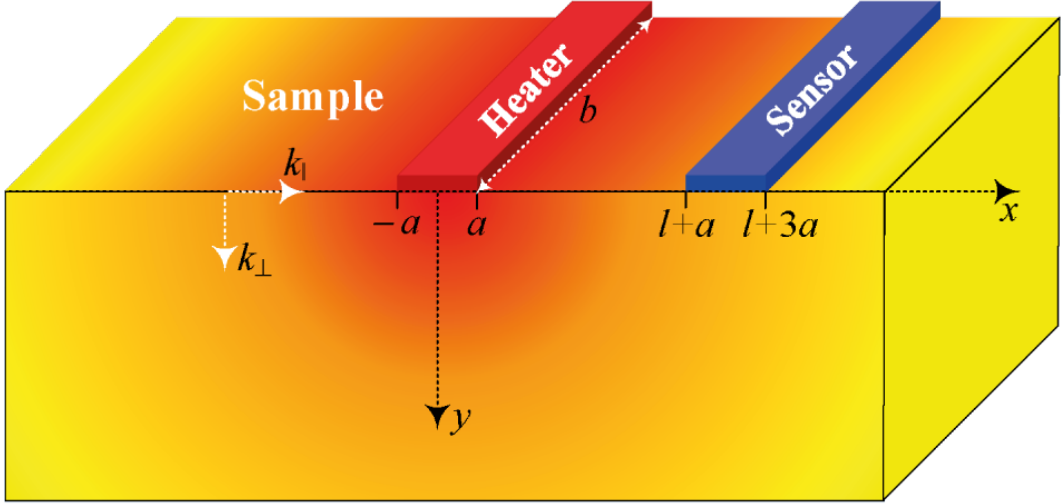

**Figure 6:** Scheme of the heater and sensor deposited on a substrate to determine its thermal properties via the 3ω/2ω method. $l$ is the separation distance between the heater and sensor pads, both of width $2a$ and length $b$.

For this thermal excitation, the linearity of the partial differential equation of heat diffusion establishes that the general temperature profile is given by

$$T(\vec{x}, t) = T_{dc}(\vec{x}) + \Re\big(T_{2\omega}(\vec{x})e^{2i\omega t}\big), \tag{1}$$

where $T_{dc}$ and $T_{2\omega}$ are the steady-state and modulated components of temperature, and $\Re(.)$ represents the real part of its argument. As a result of the temperature variation of the heater, its electrical resistance $R(T)$ varies and can conveniently be expanded around room temperature $T_0$ through the following Taylor series expansion[65]

$$R(T) = R_0[1 + \alpha(T - T_0) + \beta(T - T_0)^2 + \cdots], \tag{2}$$

where $R_0 = R(T_0)$, $\alpha = R'(T_0)/R(T_0)$, and $\beta = R''(T_0)/2R(T_0)$. Equation (2) represents the calibration curve of the heater characterized by its TCRs $\alpha, \beta, \ldots$. The Ohm's law $\Delta V = RI$ can now be used to determine the voltage difference $\Delta V$ between the extremes of the heater and sensor.

## A. Heater electric signal

According to Eq. (2), the heater voltage difference $\Delta V_h = RI_0\Re(e^{i\omega t})$ is given by

$$\Delta V_h = V_0[1 + \alpha(T - T_0) + \beta(T - T_0)^2 + \cdots]\Re(e^{i\omega t}), \tag{3}$$

where $V_0 = R_0 I_0$.

- **Linear approximation on $T - T_0$**

We here assume that $R$ and $\Delta V_h$ have a linear dependence on temperature and therefore the effect of $\beta$ and higher-order coefficients is neglected. Under this usual condition, the combination of Eqs. (1) and (3) yields

*Corresponding author: jalabert@iis.u-tokyo.ac.jp

$$\Delta V_h = \Re(V_\omega e^{i\omega t} + V_{3\omega} e^{3i\omega t}), \quad (4)$$

where the voltage components oscillating with $\omega$ and $3\omega$ are defined by

$$V_\omega = V_0 \left[1 + \alpha \left(T_{dc} + \tfrac{1}{2} T_{2\omega}\right)\right], \quad (5a)$$

$$V_{3\omega} = \tfrac{1}{2} \alpha V_0 T_{2\omega}. \quad (5b)$$

The $3\omega$ component of voltage thus determines the $2\omega$ component of temperature via Eq. (5b), provided that the TCR $\alpha$ is known. The steady-state component of temperature, on the other hand, is given by the combination of Eqs. (5a) and (5b), which establishes that $T_{dc} = \alpha^{-1}[(V_\omega - V_{3\omega})/V_0 - 1]$. The experimental measurement of $V_\omega$ and $V_{3\omega}$ hence determines both components of temperature.

- **Quadratic approximation on $T - T_0$**

We here consider the impact of the coefficients $\alpha$ and $\beta$ on the heater voltage, and neglect the effect of higher-order coefficients. By inserting Eq. (1) into Eq. (3), one obtains

$$\Delta V_h = \Re(V_\omega e^{i\omega t} + V_{3\omega} e^{3i\omega t} + V_{5\omega} e^{5i\omega t}), \quad (6)$$

where the voltage components $V_\omega$, $V_{3\omega}$ and $V_{5\omega}$ are given by

$$V_\omega = V_0 \left[1 + \alpha T_{dc} + \beta T_{dc}^2 + \left(\tfrac{\alpha}{2} + \beta T_{dc}\right) T_{2\omega} + \tfrac{\beta}{2} |T_{2\omega}|^2\right], \quad (7a)$$

$$V_{3\omega} = V_0 \left[\left(\tfrac{\alpha}{2} + \beta T_{dc}\right) T_{2\omega} + \tfrac{\beta}{4} T_{2\omega}^2\right], \quad (7b)$$

$$V_{5\omega} = \tfrac{1}{4} \beta V_0 T_{2\omega}^2, \quad (7c)$$

with $|T_{2\omega}|$ being the amplitude of $T_{2\omega}$. The fifth harmonic $V_{5\omega}$ of voltage thus determines $T_{2\omega}$, while the third ($V_{3\omega}$) and first ($V_\omega$) harmonics are related not only to $T_{2\omega}$ but also to $T_{dc}$. Considering that the propagation of thermal waves is heavily attenuated by the frequency increase, the application of the physical condition $T_{2\omega}(\omega \to \infty) = 0$ in Eq. (7a) yields

$$V_\omega(\omega \to \infty) \equiv V_\infty = V_0(1 + \alpha T_{dc} + \beta T_{dc}^2). \quad (8)$$

The high-frequency voltage $V_\infty$ therefore determines the steady-state temperature component given by the solution of Eq. (8): $\alpha + 2\beta T_{dc} = \sqrt{H}$, with $H = \alpha^2 - 4\beta(1 - V_\infty/V_0)$. The insertion of this latter relation into Eq. (7b) determines $T_{2\omega}$ in terms of $V_{3\omega}$, as follows

$$T_{2\omega} = \frac{\sqrt{H + 4\beta V_{3\omega}/V_0} - \sqrt{H}}{\beta}. \quad (9)$$

The modulated temperature component $T_{2\omega}$ (See Fig. S8a) can thus be found by recording either $V_{3\omega}$ or $V_{5\omega}$ involved in Eq. (9) or Eq. (7c), respectively. For the asymptotic case of $\beta \to 0$, Eq. (9) reduces to Eq. (5b), as expected. Equation (9) thus represents the extension of Eq. (5b) when the calibration curve $R$ vs $T$ exhibits a quadratic behavior ($\beta \neq 0$).

**B. Sensor electrical signal**

Considering that the sensor is done of the same material as the heater, they both have the same electrical resistance for a given temperature. Therefore, the circulation of a DC electrical current $I_{s0}$ along the sensor induces a voltage difference $\Delta V_s = RI_{s0}$, which according to Eq. (2) is given by

$$\Delta V_s = V_{s0}[1 + \alpha(T - T_0) + \beta(T - T_0)^2 + \cdots], \quad (10)$$

where $V_{s0} = R_0 I_{s0}$.

- **Linear approximation on $T - T_0$**

By neglecting the effect of $\beta$ and all other higher-order coefficients, the combination of Eqs. (1) and (10) yields

$$\Delta V_s = V_{dc} + \Re(V_{2\omega}e^{2i\omega t}), \quad (11)$$

where the DC and AC voltage components determine the steady-state and modulated components of temperature, as follows

$$V_{dc} = V_{s0}(1 + \alpha T_{dc}), \quad (12a)$$
$$V_{2\omega} = \alpha V_{s0} T_{2\omega}. \quad (12b)$$

The modulated components of both the electric and thermal signals of the sensor thus oscillate with the same frequency $2\omega$, while the DC voltage increases linearly with the steady-state temperature. Therefore, the experimental measurement of $V_{dc}$ and $V_{2\omega}$ determines both components of temperature. The sensor $2\omega$ voltage $V_{2\omega}$ is deduced from the measured $W_{2\omega}$ using the relation $V_{2\omega} = W_{2\omega}(1 + R_{sp}/R_2)$, where $R_{sp}$ and $R_2$ are external fixed resistances.

- **Quadratic approximation on $T - T_0$**

For an approximation up to $(T - T_0)^2$ in Eq. (10), the voltage difference across the sensor is given by the insertion of Eq. (1) into Eq. (10), which yields

$$\Delta V_s = V_m + \Re(V_{2\omega}e^{2i\omega t} + V_{4\omega}e^{4i\omega t}), \quad (13)$$

where the voltage components are defined by

$$V_m = V_{s0}\left(1 + \alpha T_{dc} + \beta T_{dc}^2 + \frac{\beta}{2}|T_{2\omega}|^2\right), \quad (14a)$$
$$V_{2\omega} = V_{s0}(\alpha + 2\beta T_{dc})T_{2\omega}, \quad (14b)$$
$$V_{4\omega} = \frac{1}{2}\beta V_{s0}T_{2\omega}^2. \quad (14c)$$

For $\beta \to 0$, Eqs. (14a) and (14b) reduce to their corresponding linear counterparts in Eqs. (12a) and (12b), as expected. For $\beta \neq 0$, on the other hand, the temperature component $T_{dc}$ can be determined by using the physical condition $T_{2\omega}(\omega \to \infty) = 0$. In this high-frequency limit ($\omega \to \infty$), Eq. (14a) takes the form

$$V_m(\omega \to \infty) \equiv V_{s\infty} = V_{s0}(1 + \alpha T_{dc} + \beta T_{dc}^2). \quad (15)$$

Equation (15) has the same mathematical form than Eq. (8) derived for the heater signal and its solution for the steady-state temperature is $\alpha + 2\beta T_{dc} = \sqrt{S}$, with $S = \alpha^2 - 4\beta(1 - V_{s\infty}/V_{s0})$. The parameter $S$ is identical to $H$ under the interchange of the voltage ratio $V_{s\infty}/V_{s0} \to V_\infty/V_0$. After inserting this expression for $T_{dc}$ into Eq. (14b), one obtains

$$T_{2\omega} = \frac{V_{2\omega}}{V_{s0}\sqrt{S}}. \quad (16)$$

The modulated temperature component (See Fig. S8b) can thus be determined not only by the fourth harmonic of the sensor voltage (Eq. (14c)), but also by its usual second harmonic through Eq. (16). Practically, $V_{4\omega} \ll V_{2\omega}$, therefore we use Eq.(14b) to calculate the sensor elevation of temperature. For the limiting case of $\beta \to 0$, Eq. (16) reduces to Eq. (12b), as expected. Therefore, Eq. (16) extends the validity of Eq. (12b) for a quadratic ($\beta \neq 0$) calibration curve $R$ vs $T$. Equations (9) and (16) thus integrate the impact of the second-order term $\beta \neq 0$ on the modulated temperature component $T_{2\omega}$ of the heater and sensor driven by the voltage components $V_{3\omega}$ and $V_{2\omega}$, respectively. The measurement of these electrical signals hence determines the thermal one, for each modulation frequency $\omega$.

## VII. MODELING OF THE THERMAL SIGNALS

*Corresponding author: jalabert@iis.u-tokyo.ac.jp

The modulated temperature components of the heater ($T_{2\omega} = \Delta T_h$) and sensor ($T_{2\omega} = \Delta T_s$) is given by the solution of the heat diffusion equation. For the anisotropic substrate shown in Fig. 6, these thermal signals are given by[51]

$$\Delta T_h = T_c \int_0^\infty \left(\frac{\sin(\xi)}{\xi}\right)^2 \frac{d\xi}{B_i + \sqrt{\xi^2 + if/f_c}}, \tag{17a}$$

$$\Delta T_s = T_c \int_0^\infty \left(\frac{\sin(\xi)}{\xi}\right)^2 \frac{\cos(\beta\xi)}{B_i + \sqrt{\xi^2 + if/f_c}} d\xi, \tag{17b}$$

where $\beta = 2 + l/a$ and $B_i = 4\varepsilon\sigma a T_0^3/\bar{\kappa}$ is the Biot number driving the radiative losses, $\varepsilon$ is the sample emissivity and $\sigma$ is the Stefan-Boltzmann constant. The characteristic temperature $T_c$ and frequency $f_c$ are related to the material properties, as follows

$$T_c = \frac{P_0}{\pi b \bar{\kappa}}, \tag{18a}$$

$$f_c = \frac{D_\parallel}{4\pi a^2}, \tag{18b}$$

with $\bar{\kappa} = \sqrt{\kappa_\parallel \kappa_\perp}$ being the geometric mean of the in-plane and cross-plane thermal conductivities of the substrate with an in-plane thermal diffusivity $D_\parallel$. Equations (17a) and (17b) apply for an infinitely thick sample for the propagation of the excited thermal waves. This condition is well satisfied by our sapphire sample, whose thickness (0.5 mm) is greater than the diffusion length, for the vast majority of frequencies considered in our experiments, as shown in the supplementary Fig. S10. For a sapphire sample with a heater width $2a = 10$ μm considered in our work, its expected thermal conductivity changes from $\bar{\kappa} = 35$ Wm$^{-1}$ K$^{-1}$ ($T_0 = 300$ K) to $\bar{\kappa} = 10$ Wm$^{-1}$ K$^{-1}$ ($T_0 = 1150$ K), and therefore the maximum ($\varepsilon = 1$) $B_i = (0.09; 17.25) \times 10^{-5}$. The Biot number for our material configuration is thus much smaller than unity, which indicates that the modulated temperature fields $\Delta T_h$ and $\Delta T_s$ are practically independent of the radiative losses. The parameters ($T_c, f_c$) and therefore the thermal properties ($\bar{\kappa}, D_\parallel$) of the sample can be determined by fitting either Eq. (17a) or Eq. (17b) to the corresponding temperature field measured as a function of the modulation frequency $f$. This fitting can accurately be done through the following analytical solution that we recently reported for the integrals in Eqs. (17a) and (17b) with $B_i = 0$[51]

$$\frac{\Delta T_h}{T_c} = \pi J(2z) + \frac{2z K_1(2z) - 1}{2z^2}, \tag{19a}$$

$$\frac{\Delta T_s}{T_c} = \frac{\pi}{8}[\beta_+^2 J(z\beta_+) + \beta_-^2 J(z\beta_-) - 2\beta^2 J(z\beta)] + \frac{1}{4z}[\beta_+ K_1(z\beta_+) + \beta_- K_1(z\beta_-) - 2\beta K_1(z\beta)], \tag{19b}$$

where $z = \sqrt{if/f_c}$, $\beta_\pm = \beta \pm 2$, and

$$J(x) = K_0(x) L_{-1}(x) + K_1(x) L_0(x), \tag{20}$$

with $K_n$ and $L_n$ being the modified Bessel function of second kind and the modified Struve function of order $n$, respectively. Eqs. (19a) and (19b) can conveniently be used to fit low- and high-frequency experimental data exhibiting a log-linear ($f \ll f_c$) and a non-log-linear ($f \approx f_c$) behavior[51]. Note that the models for both the heater and sensor temperature signals depend on the sample thermal conductivity only through the geometric mean $\bar{\kappa} = \sqrt{\kappa_\parallel \kappa_\perp}$ between the in-plane ($\kappa_\parallel$) and cross-plane ($\kappa_\perp$) components. Therefore, for the considered bulk sample, we can determine $\bar{\kappa}$ (not $\kappa_\parallel$ and $\kappa_\perp$ separately) by fitting either Eq. (19a) or (19b) to their corresponding experimental data.

## VIII. RESULTS AND DISCUSSION

*Corresponding author: jalabert@iis.u-tokyo.ac.jp

We here fit the modeled heater and sensor thermal signals in Eqs. (19a) and (19b) to their corresponding experimental counterparts to determine the mean thermal conductivity ($\bar{\kappa}$) and in-plane thermal diffusivity ($D_\parallel$) of a sapphire wafer. The measurements were recorded for modulation frequencies up to 30 kHz, which is high enough to lead the sensor signal beyond its usual log-linear regime ($f < f_c$) and therefore to gain sensitivity to $D_\parallel$[51]. Considering that the bending of the sensor signal appears for $f > f_c$, we chose a heater width sufficiently large ($2a = 10$ μm) to reduce $f_c$ in Eq. (18b) and hence facilitate the accurate measurement of $D_\parallel$, without modulating extremely high frequencies.

Figure 7(a) shows the typical fitting of the real parts of the heater and sensor temperatures recorded by our 3ω/2ω experimental setup. We consider the fitting of the real parts only because they are generally greater than the imaginary parts ($\Re(\Delta T_n) > \Im(\Delta T_n)$), exhibit the highest sensitivity to the thermal conductivity and thermal diffusivity[51], and hence they allow the accurate determination of these two thermal properties. Note that the heater signal $\Re(\Delta T_h)$ varies around unity and is one order of magnitude greater than the sensor signal $\Re(\Delta T_s)$. The heater-to-sensor heat diffusion was thus driven by a temperature difference of less than 1 K, for the vast majority of frequencies.

Considering that the radiative losses induce a low-frequency plateau on both the heater and sensor signals[51], the absence of this plateau on the measured temperature fields shown in Fig. 7(a) indicates that our measurements are insensitive to the radiative losses. This absence of significant radiative currents is explained by the fact that the 3ω signal represents the modulated temperature component, whose values are typically smaller than 1 K (See Fig. 7(a)) and therefore it is much smaller than the steady-state temperature component varying from 300 to 1150 K. The observed insensitivity of the heater and sensor signals to the radiative losses is thus expected to allow the accurate determination of the thermal conductivity and thermal diffusivity of our sample, even at high temperature.

For all the considered temperatures (300 – 1150 K), the models of $\Re(\Delta T_h)$ and $\Re(\Delta T_s)$ accurately fit (solid lines) their corresponding experimental data (solid points) for all frequencies up to 6 and 30 kHz, at which the signals nearly vanish, respectively. According to Eqs. (19a) and (19b), the log-linear behavior of $\Re(\Delta T_h)$ is highly sensitive to the thermal conductivity, while the high-frequency bending of $\Re(\Delta T_s)$ drives the values of the thermal diffusivity. Based on this fact and the experimental data shown in Fig. 7(a), we use $\Re(\Delta T_h)$ and $\Re(\Delta T_s)$ to measure $\bar{\kappa}$ and $D_\parallel$, respectively. The log-linear behavior of $\Re(\Delta T_h)$ determines the thermal conductivity while and the high-frequency bending of $\Re(\Delta T_s)$ drives the values of the in-plane thermal diffusivity shown in Fig. 7(b). Both $\bar{\kappa}$ and $D_\parallel$ monotonically decrease as the temperature increases and their values agree with those reported in the literature[49,51,66]. This remarkable agreement, within the uncertainty of 0.3% (See Table S4), confirms the capability of our experimental setup to accurately measure the thermal conductivity and thermal diffusivity of solid materials beyond the temperature limitations of previous measurements or other techniques (764 K for the reference thermal conductivity shown in Fig. 7(b)).

*Corresponding author: jalabert@iis.u-tokyo.ac.jp

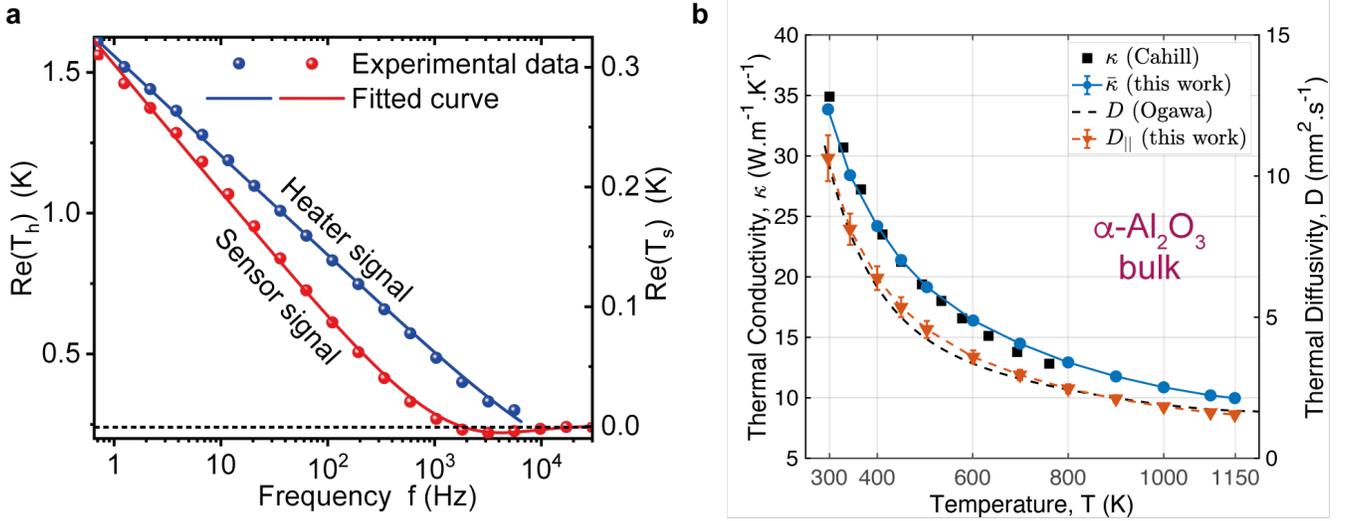

**Figure 7:** (a) Typical fitting of the real parts of the heater and sensor temperature signals obtained for a sapphire wafer and (b) its corresponding thermal conductivity ($\bar{\kappa}$) and thermal diffusivity ($D_{\parallel}$) fitted with Eqs. (19a) and (19b), respectively (Data available in Table S4). The mean values of $\bar{\kappa}$ and $D_{\parallel}$ agree with their corresponding values reported in the literature[51,66], within the error bars representing the standard deviation obtained through a least-square fit (Extended comparison with literature data is shown in Fig. S11. The thermal conductivity and diffusivity are overestimated by more than 20 % when considering only the linear TCR, as shown in Fig. S12 and Table S5).

## IX. CONCLUSIONS

We have presented an ultra-high-temperature vacuum probe station (UHT-VPS) designed for continuous electrical and thermal characterization of materials up to 1150 K. The system incorporates a radiative silicon carbide (SiC) heater that transfers heat to a physically isolated sample holder, thereby eliminating electrical leakage and avoiding the use of insulating materials that degrade at elevated temperatures. The mechanical layout accommodates thermal expansion mismatches, ensuring stable operation across a wide temperature range. A quadratic temperature coefficient of resistance (TCR) model has been implemented in conjunction with the 3ω/2ω method, enabling accurate extraction of thermal conductivity and diffusivity from a single measurement set. Experimental validation using 200 nm-thick chromium/platinum micro-resistors patterned on a bulk α-sapphire substrate shows strong agreement with reference data from 300 to 1150 K, confirming the accuracy and repeatability of the system.

The developed UHT-VPS combined with the 3ω/2ω technique can broadly applicable to bulk materials, thin films, superlattices, membranes, 2D materials, nanostructures or measurements through vacuum gaps. Its stable high-temperature performance makes it suitable for evaluating device reliability under thermal stress, anticipating the future development of ultra-high-temperature operating lifetime (UHTOL) protocols for wafer-scale testing of integrated circuits, extreme-MEMS, and power electronics. The modular design also enables future integration of a corrosion-resistant chamber and gas injection system for in-situ studies of oxidation and corrosion processes on a sample. By combining radiative heating, nonlinear thermal modeling, and high-frequency signal acquisition, the UHT-VPS provides a high-throughput platform for advanced material characterization and high-quality training dataset[67] under ultra-high temperature conditions.

**SUPPLEMENTARY MATERIAL**

*Corresponding author: jalabert@iis.u-tokyo.ac.jp

Figures S1 (absence of leakage from the SiC heater to the device under test); Figure S2 (holder temperature uniformity); Figure S3 (AFM measurements); Figure S4 (Resistances versus time); Figure S5 ($V_\omega$ versus time); Figure S6 ($W_{2\omega}$ versus time); Figure S7 (example of the last 20% of the recorded data); Figure S8 ($\Delta T_h$ and $\Delta T_s$ versus frequency); Figure S9 (overlay of linear and quadratic fits of R(T)); Figure S10 (penetration depth); Figure S11 (extended comparison of thermal properties with the literature); Figure S12 (comparison of the thermal conductivity and diffusivity obtained from the linear and quadratic fits of R(T)); Table S1 ($R_h$ and $R_s$ values); Table S2 (TCRs); Table S3 (TCR comparison with the literature for the Platinum thin film); Table S4 (thermal conductivity and diffusivity data); Table S5 (data supporting Fig. S12).

## ACKNOWLEDGEMENTS

This work was supported by the CREST JST (GrantsNo. JPMJCR19Q3 and No. JPMJCR19I1) and KAKENHI JSPS (Grants No. 21H04635 and No. JP20J13729) projects, as well as by the JSPS Core-to-Core Program (Grant No. JPJSCCA20190006).

## AUTHOR DECLARATIONS

### Conflict of Interest

The authors have no conflict to disclose.

### Author Contributions

L.J conceived the original idea of the machine, built the experimental 3ω/2ω setup, automated the data acquisition (Labview), performed the data treatment and curation (Matlab) and wrote the original manuscript. J. O-M developed the mathematical models for the 3-omega data treatment, the thermal analysis to retrieve the thermal conductivity and diffusivity, and wrote the original manuscript. Y.W, B.K and R.A fabricated the sample in clean room. M.N and S.V supervised the overall research project, secured the necessary financing and was responsible for project administration. All authors contributed to discussing the results, manuscript review and revision.

## DATA AVAILABILITY

The data that support the findings of this study are available from the corresponding author upon reasonable request.

*Corresponding author: jalabert@iis.u-tokyo.ac.jp

*Corresponding author: jalabert@iis.u-tokyo.ac.jp

*Corresponding author: jalabert@iis.u-tokyo.ac.jp

*Corresponding author: jalabert@iis.u-tokyo.ac.jp

*Corresponding author: jalabert@iis.u-tokyo.ac.jp

*Corresponding author: jalabert@iis.u-tokyo.ac.jp